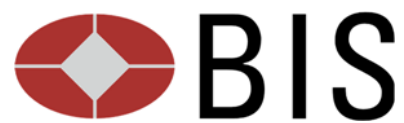

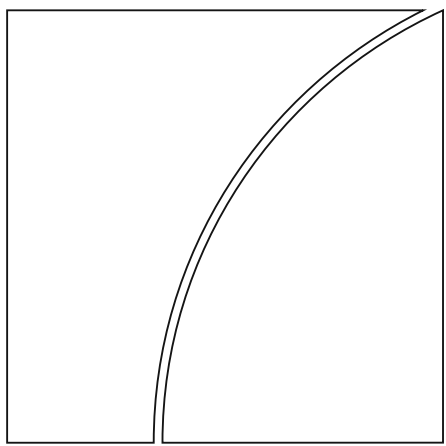

# BIS Working Papers
No 1056

# Understanding the food component of inflation

by Emanuel Kohlscheen

# Understanding the Food Component of Inflation

Emanuel Kohlscheen [1,2]

## Abstract

This article presents evidence based on a panel of 35 countries over the past 30 years that the Phillips' curve relation holds for food inflation. That is, broader economic overheating does push up the food component of the CPI in a systematic way. Further, general inflation expectations from professional forecasters clearly impact food price inflation. The analysis also quantifies the extent to which higher food production and imports, or lower food exports, reduce food inflation. Importantly, the link between domestic and global food prices is typically weak, with pass-throughs within a year ranging from 0.07 to 0.16, after exchange rate variations are taken into account.

[1] Senior Economist, Bank for International Settlements, Centralbahnplatz 2, 4002 Basel, Switzerland. *E-Mail address*: emanuel.kohlscheen@bis.org.
[2] I am grateful to Deniz Igan and Daniel Rees for helpful comments and to Emese Kuruc for research assistance. The views expressed in this paper are those of the author and do not necessarily reflect those of the Bank for International Settlements.

# 1. Introduction

Changes in food prices are often treated as exogenous shocks to macroeconomic developments, that strike according to the whims of nature. They are frequently seen as a nuisance, that is not worth modelling. Yet they are an important component of consumer baskets. Their weight on inflation outcomes is relevant in advanced and developing countries alike (see e.g. Peersman (2022)). On average, food and non-beverage items represent 17% of consumer price indices in OECD countries, ranging from a low of 8% in the United States to 27% in the case of Poland.[3] Because of the high volatility of food item prices, their contribution to the variation of inflation can easily exceed their weight. Further, because of their salience, food prices could have a disproportionate impact on inflation expectations and more readily spread to other categories.

A key research question is whether and to which extent food inflation responds to broader domestic macroeconomic conditions such as output gaps or changes in headline inflation expectations. For instance, does economic over- or underheating spill-over into food pricing? Is the Phillips relation valid for food items? Answers to these core questions are key also for policy makers, as they might for instance affect the calibration of monetary policy. Particularly in countries where the inflation target is based on headline inflation, the optimal policy is bound to differ depending on the extent to which the food component of the CPI is responsive to output gaps and inflation expectations – which are directly influenced by monetary actions. Further, how strongly do global food prices transmit to retail food costs? And by how much can larger crop yields, higher food imports – or conversely lower food exports – contribute to the alleviation of domestic food prices? This article provides quantitative answers to each of the above questions based on impulse responses obtained through the Jordà (2005) local projection method.

The cross-country analysis reveals that a Phillips' curve relation holds for food inflation. That is, broader economic overheating does push up the food component of CPIs in a systematic way. This may be partly because most retail food items incorporate some cost of services and/or packaging. Further, general inflation expectations from professional forecasters could impact food price inflation. That is, food price setters may be affected by the broader assessment regarding future inflation. Together these results highlight that food price developments are not exogenous but deeply intertwined with broader economic developments. It follows that macroeconomic stabilisation policies also affect domestic food prices (as do food policies which are much more targeted).

When it comes to the effects of domestic food production and international trade, the baseline results point out that a 10% crop yield increase reduces food CPI inflation by around 0.5%. The central estimates are that the quantitative effect of a similar increase in food imports is somewhat smaller, with price responses that are not always statistically significant, indicating imperfect substitutability when it comes to dampening the effects of eventual local crop shortfalls on the aggregate food CPI. In turn, the net effect of a 10% increase in food exports is estimated to be smaller, at 0.3%, likely owing to the concentration of exports on specific products which are not representative of the typical consumers' food consumption basket.

---

[3] See Appendix Table A1 for the weight in each country that is covered in this study.

What is noteworthy is that the connection of domestic with global food prices is typically weak, with central estimates for pass-throughs within a year ranging from 0.07 to 0.16, after exchange rate variations are considered. This underscores that global food price indices are a very imperfect proxy for domestic food price developments. Finally, the effect of energy costs on food costs, while significant from a statistical viewpoint at usual confidence levels, is found to be economically negligible.

The novelty of this article is that it fills a gap in the literature, by providing novel cross-country evidence on the factors that drive this vastly underexplored component of CPIs. While this is only a first step for a much broader research agenda, it contributes to a better understanding of retail food price pressures. Such understanding could prove key to food policies, as well as broader policies aiming at price stability.

**Relation to the literature.** As already indicated, the recent literature on the drivers of food prices from a macroeconomic perspective is surprisingly scant, particularly given the wide social and economic repercussions of the topic. While there are some studies covering specific low-income countries, studies for middle- and high-income countries are rather few. One notable recent exception is Peersman (2022)'s study, based on a SVAR-IV model for the euro area. He concludes that exogenous swings in international food prices explain a large portion of the medium-term CPI volatility. Earlier and similarly, the analysis of Ferruci et al (2010) had concluded that commodity prices were key for the 2007-08 increase in inflation.[4]

To the best of my knowledge, the link between domestic food CPI and expectations and the "food Phillips curve" uncovered in the current paper are novel. So is also the precise quantification of the effects of crop yields and international food trade to retail food prices in a broad cross-country setup. That said, country-specific studies such as Durevall et al (2013), among others, already explored the link between domestic agricultural production and food prices.

In contrast to the studies mentioned above, the direct link of domestic food prices with international prices had already been found to be on the weak side in earlier studies. Furceri et al (2016) report a 1-year pass-through from global to domestic food prices of only 0.05 for advanced economies, which is even lower than my central estimate of 0.08 for high income countries. They do conjecture that the lower pass-through in advanced economies may be linked to greater established credibility of monetary regimes in these.

Where the existing literature on food prices is most abundant is on the connection between energy and food prices. Avalos (2014) for instance looks at the increased sensitivity of corn prices to oil after biofuel promotion policies were introduced in the United States. That said, the results of the current analysis – which are based on macro data - align much better with those of Lambert and Miljkovic (2010) and Baumeister and

---

[4] Their study highlights the role of non-linearities in the pass-through from input to final prices.

Kilian (2014) who found that oil prices are not responsible for more than a negligible share of the increase in retail food prices in the US, and less with Irz et al (2013) or Baek and Who (2014) who report a non-negligible effect.[5]

## 2. Methodology

In what follows, the impact of the drivers of CPI food inflation are analysed by means of impulse responses obtained through the local projection method of Jordà (2005). Plagborg-Moller and Wolf (2021) show that this method gives results that are equivalent to the impulse responses obtained from VARs.[6] More specifically, the baseline panel model

$$\varphi_{i,t+h} = \alpha_h + \rho_{i,t}\varphi_{i,t} + \beta_{1,h}\pi_{i,t} + \beta_{2,h}E_t\pi_{i,t+1} + \beta_{3,h}\text{ygap}_{i,t} + \beta_{4,h}\Delta FY_{i,t} + \beta_{5,h}\Delta FX_{i,t} + \beta_{6,h}\Delta FM_{i,t} + \beta_{7,h}\Delta oil_t + \beta_{8,h}\varphi*_t + \beta_{9,h}trend_t + \gamma_{i,h} + \varepsilon_{i,h,t}$$

is estimated. The dependent variable $\varphi_{i,t+h}$ is the annual variation of food inflation in year $t+h$, where $h = 1$ or $h = 2$, and $\varphi_{i,t}$ stands for a lagged dependent variable. $\pi_{i,t}$ represents current inflation (that is, in country $i$ in year $t$), $E_t\pi_{i,t+1}$ the expected inflation for the following year, so that the specification has a backward-looking and a forward-looking inflation element. ygap$_{i,t}$ is the output gap, which is intended to capture any eventual Phillips type relation in food prices. $\Delta FY_{i,t}$ represents per capita crop growth in country $i$ in year $t$. $\Delta FX_{i,t}$ and $\Delta FM_{i,t}$ capture, respectively, per capita food export and import growth. $\Delta oil_t$ reflects the variation in global oil prices. $\varphi*_t$ is the second global variable, which represents variation for global food price index, as measured in domestic currency. A time trend is added in some specifications to allow for the possibility of a secular drift in food prices. $\gamma_{i,h}$ represents country fixed effects, which capture also unobserved heterogeneities, and $\varepsilon_{i,h,t}$ the error term.[7]

---

[5] The finding that the time trend in food prices is economically negligible appears to stand in contrast with support for the Prebisch-Singer hypothesis found in other studies (e.g. Baffes et al (2016) and the mixed results obtained by Arezki et al (2016)). That said, this emerges more as a by-product in our analysis, and is not the focus of the analysis. The current study is also by no means the first that finds evidence that appears to go against the hypothesis (see Gilbert (2010)).

[6] See also Olea et al (2021).

[7] Alternatively, a fertilizer price index was also included as control. This follows from the fact that fertilizer price pressures often translate into higher food prices down the road (see for instance IMF (2022, p. 38)). The fertilizer price index was constructed based on the average global price variation of phosphate, potassium and urea (unweighted). The inclusion of this additional control variable did not change the

Note that all right-hand variables in the specification are effectively lagged relative to the left-hand variable. One direct implication is that the predicted values can indeed be interpreted as in-sample forecasts of the following years' food price inflation in the respective countries and times. All economic variables are in log differences.

Food price inflation, the variable of interest, is the variation in the food CPI reported by the OECD (i.e. group 01 of the COICOP 2018 classification). This group includes basic food items,[8] non-alcoholic beverages (such as fruit juices, coffee, tea, cocoa drinks, water and soft drinks), and services for processing primary goods for food and non-alcoholic beverages.

Headline inflation numbers are sourced from the IMF's International Financial Statistics. Following year inflation expectations were taken from *Consensus Economics* for each December. The survey is based on professional forecasters responses (mostly from banks). Domestic food production and trade data are indices which were computed based on World Bank data on these variables and on the population of the respective countries. The output gap estimates are from the IMF's World Economic Outlook.[9] Oil price variations refer to Brent oil, sourced from Bloomberg, while the global food prices are based on the Food and Agriculture Organization/UN (FAO) food price index. The data frequency is annual, spanning from 1990 to 2020, for 35 countries that are listed in Appendix Table A2. The complete variable specification in Table A3 and summary statistics of these can be found in Table A4.

Figure 1 shows the sample means, ranges of variation and standard deviations of food CPIs, per capita crop, food exports and import growth for nine selected (larger) economies. Average food inflation over the 30-year period has mostly situated near to the 2% inflation targets that prevails in most countries that are depicted in the Figure. The range of variation has been largest in the cases of Australia and Sweden. Domestic crop growth has hovered around zero in per capita terms, with variation being largest in Spain.

---

coefficients of interest in any material way. As country-specific fertilizer price indices are not available, the more parsimonious specification without this component was kept as the benchmark model.

[8] This comprises cereals, live animals, fish and seafood, milk and dairy products, oils and fats, fruits and nuts, vegetables, sugar and desserts, as well as ready-made food.

[9] Vintage of April 2022.

**Figure 1 - Growth and volatility**

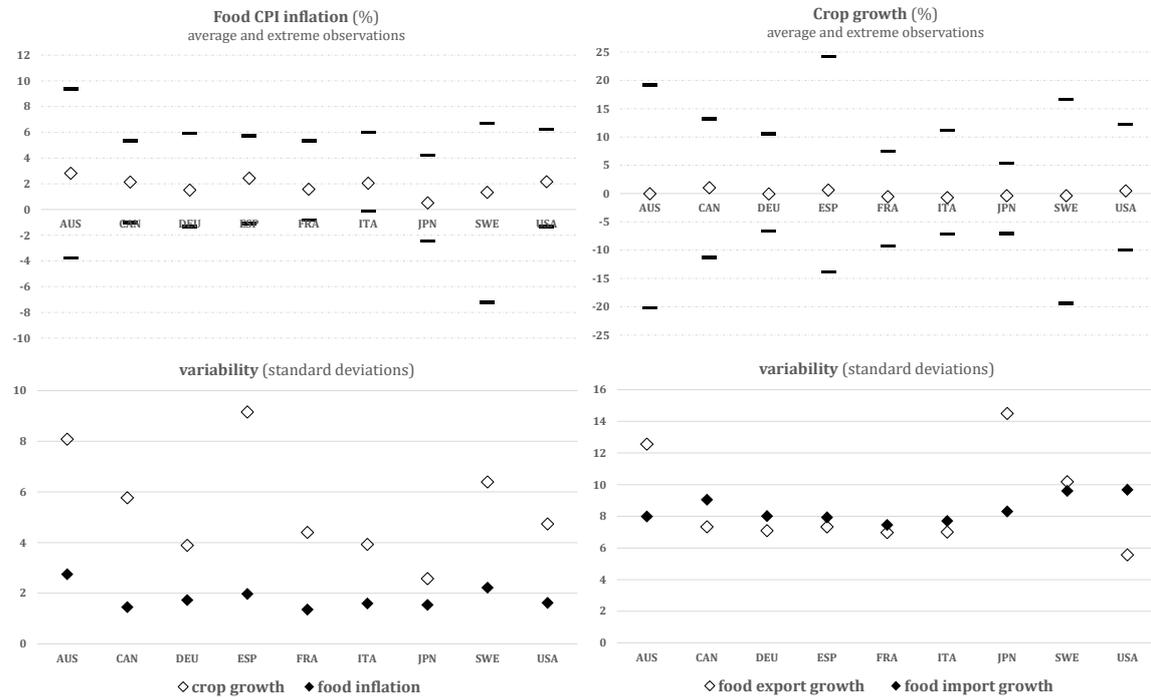

## 3. Baseline Estimates

Table 1 shows the baseline estimation results for the Jordà (2005) methodology for years *t+1* and *t+2*, without and with a time trend. Overall, the model explains a large fraction of the variation in Food CPI inflation. For one year ahead inflation, the within R2 is 0.61, indicating that the explanatory power comes also from the time series dimension.[10]

    The coefficients in the first four rows indicate that broad macroeconomic conditions do impact retail food inflation in a very significant way. First, food price increases in one year partly do spill over into the following year, as indicated by the persistence coefficient in the lagged dependent variable. Second, when it comes to general CPI inflation, there appears to be some mean reversion which spills over onto food prices. Yet, most notably, retail food prices are clearly affected by overall inflation expectations. Further, the output gap measure for the broader economy does translate into 1-year forward inflation, with a coefficient (0.12) that is comparable in magnitude to those obtained for the aggregate CPI (see e.g. Stock and Watson (2019) and Hazell et

---

[10] Without the lagged dependent variable, the within R2 is still 0.59, with coefficients for other variables that are qualitatively similar. This indicates that the high explanatory power of the model does not come from the lagged dependent variable.

al (2022)). Together, these results suggest that retail food inflation is not independent of core economic developments. On the contrary, it hinges on variables that are affected by monetary policy. [11]

### Table 1 – Baseline Estimation Results

Dependent variable: Food CPI – 1 and 2 year forward

|  | 1 year forward | 2 year forward | 1 year forward | 2 year forward |
|---|---|---|---|---|
| lagged dep. var. | 0.2226*** | -0.0565 | 0.1924*** | -0.0754 |
|  | 0.0462 | 0.0627 | 0.0403 | 0.0594 |
| CPI inflation | -0.3015*** | -0.1934* | -0.2317*** | -0.1471 |
|  | 0.0813 | 0.1061 | 0.0744 | 0.1019 |
| expected inflation | 1.0865*** | 1.1109*** | 1.1279*** | 1.1358*** |
|  | 0.0971 | 0.0695 | 0.1047 | 0.0700 |
| output gap | 0.1161** | 0.0499 | 0.1169** | 0.0538 |
|  | 0.0475 | 0.0421 | 0.0441 | 0.0417 |
| domestic crop growth | -0.0471*** | -0.0153 | -0.0471*** | -0.0140 |
|  | 0.0115 | 0.0159 | 0.0112 | 0.0158 |
| food exports growth | 0.0300** | -0.0016 | 0.0295** | -0.0016 |
|  | 0.0119 | 0.0147 | 0.0115 | 0.0143 |
| food imports growth | -0.0151 | -0.0273 | -0.0111 | -0.0261 |
|  | 0.0160 | 0.0165 | 0.0155 | 0.0164 |
| oil price change | 0.0131*** | 0.0076*** | 0.0132*** | 0.0079*** |
|  | 0.0023 | 0.0025 | 0.0023 | 0.0025 |
| global food price inflation (FAO) | 0.1105*** | -0.0097 | 0.1112*** | -0.0100 |
|  | 0.0131 | 0.0146 | 0.0131 | 0.0142 |
| time trend |  |  | 0.0004*** | 0.0003* |
|  |  |  | 0.0001 | 0.0002 |
| observations | 832 | 819 | 832 | 819 |
| number of countries | 35 | 35 | 35 | 35 |
| country fixed effects | yes | yes | yes | yes |
| R2 between | 0.961 | 0.924 | 0.967 | 0.940 |
| R2 within | 0.606 | 0.340 | 0.614 | 0.343 |
| RMSE | 0.019 | 0.024 | 0.019 | 0.024 |

Note: Explanatory variables at year $t$, and dependent variables in year $t+1$ and $t+2$. All variables in log changes. Estimated on yearly data. Cluster-robust standard errors are shown below coefficients. \*\*\*/\*\*/\* denote statistical significance at 1/5/10% confidence level.

---

[11] Note that this does not imply that central banks should target an inflation measure that includes food inflation. Even after the effect of monetary policy, food inflation is still more volatile than headline inflation. Targeting a more stable measure of inflation is desirable in many instances.

Further, the coefficients in the next three rows show that domestic crops and international food trade impinge on following years' retail food prices. The coefficient is such that a 10% crop increase has the effect of reducing next years' food price by about 47 basis points. The central estimates indicate that food imports growth have a somewhat smaller effect, which is not statistically significant, attesting to at best partial substitutability between own and imported production. By contrast, a 10% increase in food exports, boosts retail food prices by only 0.30 percentage points (with *p*-values < 0.05). The smaller (absolute) magnitude of the later effect is likely due to the imperfect substitutability between food exports, which tend to be concentrated on a handful of products, and the typical domestic food consumption basket.

Third, the effect of global commodity prices on domestic retail food prices is found to be very modest. Particularly oil price changes, have very small quantitative effects on retail food inflation. In turn, the pass-through of global food prices, as measures by the FAO food price index, is very partial, with a coefficient of only 0.11 within a year. This can be rationalized by the high segmentation in the food market, particularly for quickly perishable products.

Finally, the time trend, when included attains a significant positive sign indicating rising nominal prices over time, all else equal. *Prima facie*, this would appear to go counter to the Prebisch-Singer hypothesis of falling commodity prices as incomes rise (see e.g. Arezki et al (2016) and Baffes et al (2016)). That said, the coefficient is almost negligible: it would take around 25 years for it to lead to a one percentage point increase in food prices, *ceteris paribus*.

As a robustness check, the estimations were repeated using Driscoll-Kraay (1998) standard errors, that factor in eventual cross-country dependence. The results, which are shown in Table A5, indicate that the conclusions are barely affected by this change in estimation method.

## 4. Results by Country Groupings

Going further, results by income level are compared. For this, the countries listed in Appendix Table A2, were grouped into (relatively) higher and lower income countries. Higher income countries are those whose nominal GDP per capita at the end of the sample period exceeded $ 30.000, according to International Monetary Fund statistics. The remaining countries were grouped as relatively lower income countries. The cut-off level of $ 30.000 corresponds to the current per capita income levels of Slovenia and Spain (which are thus included in the relatively higher income group). Note that as the majority of countries in the full sample are relatively well off (all being OECD members), the lower income group has a smaller number of observations. This tends to increase the uncertainty around the magnitude of the coefficients in this group.

## Table 2 – Results by Country Income Level

Dependent variable: Food CPI – 1 and 2 year forward

|  | higher income countries | | lower income countries | |
| --- | --- | --- | --- | --- |
|  | 1 year forward | 2 year forward | 1 year forward | 2 year forward |
| lagged dep. var. | 0.1947*** | -0.0224 | 0.1869** | -0.1630 |
|  | 0.054 | 0.055 | 0.079 | 0.138 |
| CPI inflation | -0.3890*** | -0.3923*** | -0.2664** | -0.0353 |
|  | 0.1226 | 0.1331 | 0.1156 | 0.1730 |
| expected inflation | 0.8122*** | 0.8896*** | 1.1297*** | 1.1476*** |
|  | 0.1452 | 0.1246 | 0.1247 | 0.0850 |
| output gap | 0.2345*** | 0.1552*** | 0.0417 | -0.0138 |
|  | 0.0625 | 0.0520 | 0.0503 | 0.0586 |
| domestic crop growth | -0.0475*** | -0.0088 | -0.0511*** | -0.0330 |
|  | 0.0143 | 0.0138 | 0.0138 | 0.0358 |
| food exports growth | 0.0177 | -0.0081 | 0.0552** | 0.0124 |
|  | 0.0131 | 0.0178 | 0.0197 | 0.0230 |
| food imports growth | -0.0347 | -0.0145 | 0.0014 | -0.0290 |
|  | 0.0243 | 0.0183 | 0.0150 | 0.0227 |
| oil price change | 0.0099*** | 0.0102*** | 0.0172*** | 0.0014 |
|  | 0.0027 | 0.0025 | 0.0052 | 0.0049 |
| global food price inflation (FAO) | 0.0776*** | 0.0032 | 0.1583*** | -0.0119 |
|  | 0.0149 | 0.0159 | 0.0129 | 0.0274 |
| observations | 582 | 577 | 250 | 242 |
| number of countries | 22 | 22 | 13 | 13 |
| country fixed effects | yes | yes | yes | yes |
| R2 between | 0.703 | 0.344 | 0.977 | 0.938 |
| R2 within | 0.397 | 0.205 | 0.765 | 0.450 |
| RMSE | 0.017 | 0.019 | 0.022 | 0.033 |

Note: Explanatory variables at year $t$, and dependent variables in year $t+1$ and $t+2$. All variables in log changes. Estimated on yearly data. Cluster-robust standard errors are shown below coefficients. ***/**/* denote statistical significance at 1/5/10% confidence level.

The estimates shown in Table 2 highlight that the positive effect of the output gap on food CPI inflation is stronger and statistically significant in the higher income group.[12] It also lasts longer, reaching not only one-year ahead food inflation, but also the two-year ahead inflation. Further, if anything the effects of international food trade on the food CPI are stronger in the lower income group. In this group, the link to global food prices is also slightly more visible, even if it remains modest (i.e. pass-through of 0.16 within a year).

Here it is important to note that, in general, a modest coefficient for the output gap variable can itself be a natural result of a successful inflation control policy, which firmly anchors inflation expectations around its target. Such firmer anchoring, will naturally lead to prices reacting less to output dislocations than would be the case in a regime where price expectations are unanchored (see Stock and Watson (2019) for more on this mechanism).

Second, as a further dissection of the data, the sample of the previous section is divided according to economic relevance of the agricultural sector in the respective countries. Namely, countries in which the sample average weight of the agricultural sector in GDP is above the overall sample mean for that variable are labelled as countries with a more relevant agricultural sector. The estimation results in Table 3 show that the sensitivity of prices to food trade is somewhat larger in countries where the weight of the agricultural sector is above the median.

Third, the sample is split into countries where the dependence on food imports is below and above the median. Estimation results in Table 4 shows that the sensitivity of food prices to trade is larger in the group of countries that rely more heavily on imports.

---

[12] Average inflation in the higher income countries over the sample period was 2.0%, while that in the lower income group was 4.8%.

# Table 3 – Results by Size of Agricultural Sector

Dependent variable: Food CPI – 1 and 2 year forward

|  | countries with more relevant agricultural sector | | countries with less relevant agricultural sector | |
| --- | --- | --- | --- | --- |
|  | 1 year forward | 2 year forward | 1 year forward | 2 year forward |
| lagged dep. var. | 0.1375** | -0.1872* | 0.2498*** | -0.0090 |
|  | 0.0609 | 0.0924 | 0.0562 | 0.0724 |
| CPI inflation | -0.1979* | 0.1009 | -0.3441** | -0.4292*** |
|  | 0.1004 | 0.1252 | 0.1384 | 0.1390 |
| expected inflation | 1.0542*** | 1.1350*** | 0.9037*** | 1.3670*** |
|  | 0.1049 | 0.0691 | 0.1820 | 0.2675 |
| output gap | 0.1194* | 0.0640 | 0.1285** | 0.1020 |
|  | 0.0608 | 0.0522 | 0.0451 | 0.0632 |
| domestic crop growth | -0.0599*** | -0.0050 | -0.0475*** | -0.0331 |
|  | 0.0159 | 0.0240 | 0.0118 | 0.0195 |
| food exports growth | 0.0370** | 0.0231 | 0.0018 | -0.0083 |
|  | 0.0174 | 0.0183 | 0.0135 | 0.0175 |
| food imports growth | -0.0108 | -0.0371* | -0.0156 | -0.0290 |
|  | 0.0212 | 0.0207 | 0.0248 | 0.0201 |
| oil price change | 0.0139*** | 0.0059 | 0.0117*** | 0.0057* |
|  | 0.0034 | 0.0043 | 0.0034 | 0.0030 |
| global food price inflation (FAO) | 0.1369*** | -0.0344 | 0.0716*** | -0.0074 |
|  | 0.0172 | 0.0216 | 0.0202 | 0.0193 |
| observations | 382 | 364 | 427 | 410 |
| number of countries | 18 | 18 | 17 | 17 |
| country fixed effects | yes | yes | yes | yes |
| R2 between | 0.965 | 0.923 | 0.708 | 0.465 |
| R2 within | 0.680 | 0.427 | 0.375 | 0.237 |
| RMSE | 0.022 | 0.029 | 0.015 | 0.017 |

Note: Explanatory variables at year $t$, and dependent variables in year $t+1$ and $t+2$. All variables in log changes. Estimated on yearly data. Cluster-robust standard errors are shown below coefficients. ***/**/* denote statistical significance at 1/5/10% confidence level.

**Table 4 – Results by Relevance of Food Imports**

Dependent variable: Food CPI – 1 and 2 year forward

|  | countries where food imports are less relevant | | countries where food imports are more relevant | |
|---|---|---|---|---|
|  | 1 year forward | 2 year forward | 1 year forward | 2 year forward |
| lagged dep. var. | 0.103* | -0.139 | 0.324*** | -0.102 |
|  | 0.053 | 0.093 | 0.052 | 0.104 |
| CPI inflation | -0.203 | 0.139 | -0.355*** | -0.149 |
|  | 0.159 | 0.216 | 0.054 | 0.155 |
| expected inflation | 1.033*** | 1.204*** | 1.122*** | 1.189*** |
|  | 0.208 | 0.167 | 0.059 | 0.042 |
| output gap | 0.192*** | 0.054 | 0.045 | 0.069 |
|  | 0.051 | 0.083 | 0.059 | 0.044 |
| domestic crop growth | -0.061*** | -0.016 | -0.050*** | -0.001 |
|  | 0.018 | 0.024 | 0.010 | 0.018 |
| food exports growth | 0.024 | -0.002 | 0.026** | 0.018 |
|  | 0.022 | 0.022 | 0.009 | 0.019 |
| food imports growth | 0.006 | -0.037 | -0.046* | -0.034 |
|  | 0.021 | 0.024 | 0.024 | 0.021 |
| oil price change | 0.014*** | 0.006 | 0.012*** | 0.006* |
|  | 0.004 | 0.004 | 0.002 | 0.003 |
| global food price inflation (FAO) | 0.129*** | -0.045* | 0.079*** | -0.008 |
|  | 0.017 | 0.023 | 0.021 | 0.021 |
| observations | 382 | 364 | 427 | 410 |
| number of countries | 18 | 18 | 17 | 17 |
| country fixed effects | yes | yes | yes | yes |
| R2 between | 0.945 | 0.901 | 0.975 | 0.935 |
| R2 within | 0.528 | 0.326 | 0.712 | 0.438 |
| RMSE | 0.022 | 0.026 | 0.015 | 0.022 |

Note: Explanatory variables at year $t$, and dependent variables in year $t+1$ and $t+2$. All variables in log changes. Estimated on yearly data. Cluster-robust standard errors are shown below coefficients. \*\*\*/\*\*/\* denote statistical significance at 1/5/10% confidence level.

## 5. Results by Quantiles and with Winsorized Variables

To further assess the robustness of the baseline results, the sensitivity of retail prices to the different explanatory factors was checked for different levels of food price increases. For this, the quantile regression approach was used, following Koenker and Bassett (1978), Koenker and Hallock (2001) and Koenker (2005). The main advantage of this method over sample partitioning, is that here the full set of observations is used in the estimation of coefficients for each quantile. This aids more precise estimation. Standard errors were obtained by bootstrapping, with 1,000 replications.

By and large, the patterns that were reported before hold for the 10th, 30th, 50th, 70th and 90th regression quantiles – as can be attested by the respective coefficients in Table 5, and their economic and statistical significance. For some factors, the sensitivity increases at the higher quantiles of the dependent variable. For instance, the effect of food exports on retail prices is somewhat larger when these are increasing at a faster rate. Similarly, the sensitivity to global food prices tends to increase in these times, even if it is quite modest overall.

Finally, to check for eventual sensitivity of the baseline results to outlier observations, the estimations in Table 1 are repeated using only winsorized (dependent and explanatory) variables. That is, observations that fell outside the 2nd / 98th percentile

## Table 5 – Quantile regressions (1 year forward)

Dependent variable: Food CPI – 1 year forward

|  | quantile | | | | |
| --- | --- | --- | --- | --- | --- |
|  | 10th | 30th | 50th | 70th | 90th |
| lagged dep. var. | 0.2046*** | 0.2609*** | 0.2641*** | 0.3357*** | 0.3561*** |
|  | 0.0600 | 0.0470 | 0.0470 | 0.0460 | 0.0730 |
| CPI inflation | -0.3232** | -0.2958*** | -0.2641*** | -0.3336*** | -0.4126*** |
|  | 0.1310 | 0.1060 | 0.0690 | 0.0630 | 0.1140 |
| expected inflation | 1.0139*** | 0.9694*** | 1.0490*** | 1.1349*** | 1.2780*** |
|  | 0.1680 | 0.1180 | 0.0880 | 0.0720 | 0.1050 |
| output gap | 0.0156 | 0.1067*** | 0.0837*** | 0.0993** | 0.0930** |
|  | 0.0520 | 0.0290 | 0.0280 | 0.0450 | 0.0470 |
| domestic crop growth | -0.0450* | -0.0480*** | -0.0454*** | -0.0501** | -0.0406** |
|  | 0.0260 | 0.0150 | 0.0120 | 0.0120 | 0.0190 |
| food exports growth | 0.0117 | 0.0105 | 0.0257** | 0.0301** | 0.0435*** |
|  | 0.0160 | 0.0130 | 0.0130 | 0.0140 | 0.0140 |
| food imports growth | -0.0205 | -0.0170 | -0.0213 | -0.0289 | -0.0330 |
|  | 0.0220 | 0.0140 | 0.0160 | 0.0230 | 0.0370 |
| oil price change | 0.0106*** | 0.0113*** | 0.0079*** | 0.0080*** | 0.0120*** |
|  | 0.0030 | 0.0020 | 0.0020 | 0.0020 | 0.0030 |
| global food price inflation (FAO) | 0.0649*** | 0.0803*** | 0.0836*** | 0.0976*** | 0.1247*** |
|  | 0.0190 | 0.0100 | 0.0120 | 0.0160 | 0.0240 |
| observations | 854 | 854 | 854 | 854 | 854 |
| pseudo R2 | 0.260 | 0.346 | 0.412 | 0.483 | 0.602 |

Note: ***/**/* denote statistical significance at 1/5/10% confidence level.

of each variable were assigned the value of the 2nd / 98th percentile. The local projection estimations after this transformation are shown in Table 6. As it turns out, the changes in coefficients are generally modest. Thus, the food CPI Phillips curve that this paper uncovers, as well as the other results are not driven by outlier observations.

## Table 6 – With winsorized variables (with $p$ = 0.02)

Dependent variable: Food CPI – 1 and 2 year forward

|  | 1 year forward | 2 year forward | 1 year forward | 2 year forward |
|---|---|---|---|---|
| lagged dep. var. | 0.2077*** | -0.0520 | 0.1779*** | -0.0779 |
|  | 0.0511 | 0.0611 | 0.0467 | 0.0606 |
| CPI inflation | -0.2586*** | -0.1755 | -0.1988*** | -0.1236 |
|  | 0.0776 | 0.0922 | 0.0711 | 0.0932 |
| expected inflation | 1.1340*** | 1.1967*** | 1.2487*** | 1.2962*** |
|  | 0.1176 | 0.0949 | 0.1275 | 0.1001 |
| output gap | 0.1203** | 0.0517 | 0.1203** | 0.0517 |
|  | 0.0476 | 0.0505 | 0.0443 | 0.0477 |
| domestic crop growth | -0.0473*** | -0.0191 | -0.0457*** | -0.0178 |
|  | 0.0137 | 0.0157 | 0.0132 | 0.0156 |
| food exports growth | 0.0229* | 0.0010 | 0.0237* | 0.0016 |
|  | 0.0121 | 0.0117 | 0.0120 | 0.0114 |
| food imports growth | -0.0029 | -0.0210 | -0.0005 | -0.0189 |
|  | 0.0151 | 0.0159 | 0.0149 | 0.0159 |
| oil price change | 0.0122*** | 0.0086*** | 0.0121*** | 0.0085*** |
|  | 0.0021 | 0.0022 | 0.0021 | 0.0022 |
| global food price inflation (FAO) | 0.1103*** | -0.0022 | 0.1106*** | -0.0019 |
|  | 0.0108 | 0.0140 | 0.0112 | 0.0137 |
| time trend |  |  | 0.0005*** | 0.0004*** |
|  |  |  | 0.0001 | 0.0001 |
| observations | 854 | 854 | 854 | 854 |
| number of countries | 35 | 35 | 35 | 35 |
| country fixed effects | yes | yes | yes | yes |
| R2 between | 0.963 | 0.922 | 0.970 | 0.941 |
| R2 within | 0.556 | 0.295 | 0.568 | 0.305 |
| RMSE | 0.018 | 0.022 | 0.018 | 0.021 |

Note: Observations for all variables are winsorized at the 2nd and 98th percentiles. Explanatory variables at year *t*, and dependent variables in year *t+1* and *t+2*. All variables in log changes. Estimated on yearly data. Cluster-robust standard errors are shown below coefficients. ***/**/* denote statistical significance at 1/5/10% confidence level.

# 6. Concluding Remarks

This article has examined the relation of domestic food CPI inflation with macroeconomic factors, food production, international trade and global food and energy prices based on impulse responses obtained through local projections which were based on a large number of countries, thus filling an important gap in the literature.

The results highlight that, rather than being purely exogenous, food CPI inflation is linked to the broader macroeconomic developments. It hinges on general consumer price expectations. Further, the study uncovers a Phillips curve relation for retail food that is very similar to that found for the aggregate CPIs (Stock and Watson (2019) and Hazell et al (2022)).

Domestic crop yields are found to be key for retail food prices. *Ceteris paribus*, a 10% higher crop means that 47 basis points are taken off food CPI inflation. The sensitivity of prices to international food trade volumes indicates only imperfect substitutability between the domestic and foreign sources of food supply. Still, variations in food trade may have a helpful dampening effect on retail food price variation when weather shocks are not perfectly correlated across countries.

Further research may want to explore the above relations using more granular price data. Another interesting extension would be to examine the Phillips curve relation for food based on more granular location data. This is left for future research efforts.

# Appendix

## Table A1 - Food weight in CPI basket by country

| | 2000 | 2021 |
|---|---|---|
| United States | 9.56 | 7.63 |
| United Kingdom | 11.40 | 9.20 |
| Germany | 10.34 | 9.69 |
| Austria | 13.61 | 11.35 |
| Australia | 13.80 | 11.39 |
| Ireland | 14.09 | 11.42 |
| Canada | 11.62 | 11.83 |
| Netherlands | 11.08 | 12.43 |
| Denmark | 14.62 | 12.58 |
| Switzerland | 11.51 | 12.61 |
| Luxembourg | 14.44 | 12.75 |
| Norway | 11.91 | 12.86 |
| Finland | 15.76 | 13.27 |
| Israel | 15.30 | 13.49 |
| Sweden | 13.40 | 14.30 |
| Iceland | 16.85 | 14.63 |
| France | 15.61 | 14.68 |
| New Zealand | 14.75 | 14.89 |
| Colombia | 23.13 | 15.05 |
| Korea | 15.84 | 15.45 |
| Slovenia | 21.84 | 16.56 |
| Belgium | 19.20 | 17.34 |
| Czech Republic | 19.76 | 17.81 |
| Italy | 16.28 | 18.37 |
| Chile | 18.90 | 19.30 |
| Slovak Republic | 29.18 | 20.61 |
| Greece | 18.49 | 21.06 |
| Hungary | 20.49 | 21.59 |
| Portugal | 22.67 | 21.95 |
| Lithuania | 36.65 | 21.97 |
| Estonia | 26.95 | 22.33 |
| Spain | 21.51 | 22.61 |
| Türkiye | 29.42 | 25.32 |
| Latvia | 32.10 | 25.52 |
| Mexico | 27.09 | 25.76 |
| Poland | 30.10 | 26.59 |
| average | 18.59 | 16.56 |

Note: This table reports the total weight of food and non alcoholic beverage in the CPI basket. Ireland data for the 2000 column is based on data for 2001, Canada for 2014, Chile for 2012 and Turkey for 2003. Source: OECD.

## Table A2 – Countries in panel

Australia, Austria, Belgium, Canada, Chile, Colombia, Costa Rica, Czechia, Denmark, Estonia, Finland, France, Germany, Greece, Hungary, Ireland, Italy, Japan, Korea, Latvia, Lithuania, Mexico, Netherlands, New Zealand, Norway, Poland, Portugal, Slovakia, Slovenia, Spain, Sweden, Switzerland, Turkey, United Kingdom and United States.

**Table A3 – List of variables**

| variable | | source |
|---|---|---|
| Food CPI inflation | yearly variation of Food CPI index (food and non-alcoholic beverages) | OECD |
| CPI inflation | yearly variation of CPI index | IMF International Financial Statistics |
| expected inflation | professional forecasters mean expected inflation for following year, surveyed in December | Consensus Economics |
| output gap | relative to potential output | IMF World Economic Outlook, April 2022 |
| domestic crop per capita growth | yearly variation of ratio of domestic crop production index to population | World Bank Development Indicators |
| food exports per capita growth | yearly variation of ratio of food imports to population | World Bank Development Indicators |
| food imports per capita growth | yearly variation of ratio of food exports to population | World Bank Development Indicators |
| population | | World Bank Development Indicators |
| oil price change | Brent barrel price in USD | Bloomberg |
| global food price inflation | index | FAO, United Nations |

**Table A4 – Summary statistics**

| | mean | std. dev. | min | max |
|---|---|---|---|---|
| CPI inflation | 2.82 | 3.72 | -8.24 | 35.24 |
| expected inflation (Consensus) | 2.65 | 2.41 | -2.62 | 28.36 |
| output gap | -0.25 | 3.04 | -16.66 | 19.79 |
| domestic crop per capita growth | 0.16 | 5.56 | -23.80 | 24.26 |
| food exports per capita growth | 3.07 | 10.47 | -80.71 | 98.08 |
| food imports per capita growth | 3.13 | 8.72 | -52.32 | 41.53 |
| oil price change | 3.58 | 36.16 | -80.48 | 83.81 |
| global food price inflation (FAO) | 2.87 | 10.54 | -34.74 | 77.83 |

## Table A5 - Baseline Results using Driscoll-Kraay Standard Errors

Dependent variable: Food CPI – 1 and 2 year forward

|  | 1 year forward | 2 year forward | 1 year forward | 2 year forward |
|---|---|---|---|---|
| lagged dep. var. | 0.2355*** | -0.0378 | 0.2087*** | -0.0577 |
|  | 0.0696 | 0.0660 | 0.0579 | 0.0650 |
| CPI inflation | -0.3162*** | -0.2100 | -0.2529*** | -0.1628 |
|  | 0.0809 | 0.1384 | 0.0622 | 0.1298 |
| expected inflation | 1.0697*** | 1.1144*** | 1.1070*** | 1.1422*** |
|  | 0.0583 | 0.0904 | 0.0651 | 0.0919 |
| output gap | 0.1161*** | 0.0448 | 0.1188*** | 0.0468 |
|  | 0.0437 | 0.0550 | 0.0435 | 0.0555 |
| domestic crop growth | -0.0479*** | -0.0180 | -0.0471*** | -0.0174 |
|  | 0.0135 | 0.0182 | 0.0131 | 0.0182 |
| food exports growth | 0.0309*** | -0.0011 | 0.0310*** | -0.0010 |
|  | 0.0102 | 0.0195 | 0.0101 | 0.0193 |
| food imports growth | -0.0181 | -0.0282 | -0.0155 | -0.0263 |
|  | 0.0189 | 0.0238 | 0.0181 | 0.0237 |
| oil price change | 0.0129* | 0.0079* | 0.0130* | 0.0080** |
|  | 0.0066 | 0.0039 | 0.0065 | 0.0038 |
| global food price inflation (FAO) | 0.1095*** | -0.0098 | 0.1093*** | -0.0100 |
|  | 0.0224 | 0.0172 | 0.0220 | 0.0166 |
| time trend |  |  | 0.0004* | 0.0003* |
|  |  |  | 0.0002 | 0.0002 |
| observations | 854 | 854 | 854 | 854 |
| number of countries | 35 | 35 | 35 | 35 |
| country fixed effects | yes | yes | yes | yes |
| R2 within | 0.617 | 0.337 | 0.624 | 0.340 |

Note: Explanatory variables at year $t$, and dependent variables in year $t+1$ and $t+2$. All variables in log changes. Estimated on yearly data. Driscoll and Kraay (1998) standard errors are shown below coefficients. ***/**/* denote statistical significance at 1/5/10% confidence level.

# Previous volumes in this series

All volumes are available on our website www.bis.org.